\documentclass[aip,
jap,
numerical,
preprint,
showpacs,
amsmath,
amssymb]{revtex4-1}
\usepackage{graphicx}
\usepackage{dcolumn}
\usepackage{bm}
\usepackage{textcomp}
\usepackage{revsymb}

\begin{document}


\title{An initial phase of Ge hut array formation at low temperature on Si(001) }

\author{Larisa V. Arapkina}
\email{arapkina@kapella.gpi.ru}
\noaffiliation

\author{Vladimir A. Yuryev}
\email{vyuryev@kapella.gpi.ru} 
\homepage{http://www.gpi.ru/eng/staff\_s.php?eng=1\&id=125}
\noaffiliation

\affiliation{A.\,M.\,Prokhorov General Physics Institute of the Russian Academy of Sciences, 38 Vavilov Street, Moscow, 119991, Russia}

\date{\today}%

\begin{abstract}
We report a direct STM observation of Ge hut array nucleation on the Si(001) surface during ultrahigh vacuum molecular-beam epitaxy at 360\,\textcelsius. Nuclei of pyramids and wedges have been observed on the  wetting layer 
$M\times N$ patches starting from the coverage of about $5.1$\,\r{A} ($\sim 3.6$\,ML). Further development of hut arrays  consists in simultaneous growth of the formerly appeared clusters and nucleation of new ones resulting in gradual rise of  hut number  density with increasing surface coverage. Huts nucleate reconstructing the patch surface from the usual $c(4\times 2)$ or $p(2\times 2)$ structure to one of two recently described  formations composed by epitaxially oriented Ge dimer pairs and  chains of four dimers.

\end{abstract}

\maketitle

\section{Introduction}

Dense  arrays of small self-assembled Ge/Si(001) clusters faceted by the $\{105\}$ planes and coherent with the substrate lattice, known as ``hut'' clusters,\cite{Mo,LeGoues_Copel_Tromp,Eaglesham_Cerullo} which are usually obtained by ultrahigh vacuum molecular beam epitaxy (UHV MBE) at lowered temperatures ($\alt 500$\,\textcelsius), has been the subject of numerous  investigations recently, mainly because of their potential  applicability in  optoelectronic devices monolithically integrated into Si chips, first of all in photosensitive structures of quantum dot infrared photodetector (QDIP) arrays.\cite{[{See, e.\,g., }] Wang-Cha,*Dvur-IR-20mcm} However, in spite of their technological importance and obvious  attractiveness of investigation of physical processes resulting in their formation, almost nothing is known about how they grow and very little is known about how they nucleate.

\begin{figure}[h]
\includegraphics[scale=0.8 ]{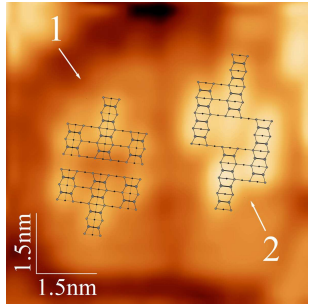}
\caption{\label{fig:nuclei}(Color online)  
STM images of pyramid (1) and wedge (2) nuclei arisen on the adjacent  $M\times N$ patches  of Ge wetting layer ($h_{\rm Ge}=6.0$~\r{A}, $U_{\rm s}=+2.60$~V, $I_{\rm t}=80$~pA); the structural models\cite{Hut_nucleation,CMOS-compatible-EMRS} are superimposed on the corresponding images.}
\end{figure}

Since their discovery by Mo et al.,\cite{Mo} it has been known that deposition of Ge on Si(001) beyond 3\,ML (1\,ML\,$\approx$\,1.4\,\AA) leads to formation of huts\cite{Mo,LeGoues_Copel_Tromp,Chem_Rev} on wetting layer (WL) with high number density ($\agt 10^{10}$\,cm$^{-2}$, Refs.~\onlinecite{Jesson_Kastner_Voigt,classification,CMOS-compatible-EMRS}). Some later the value of Ge coverage, at which 3D clusters emerged, was confirmed by Iwawaki et al.\cite{Iwawaki_initial} who, in the course of a comprehensive STM study of the low-temperature epitaxial growth of Ge on Si(001),\cite{Iwawaki_initial,Iwawaki_SSL,*Iwawaki_dimers,*Nishikawa_105}  directly observed appearance of minute (a few ML high) 3D Ge islands at 300{\,\textcelsius} on  $(M\times N)$-patched  WL; deposition of 4\,ML of Ge resulted in formation of a dense array of small huts. 
Various values of Ge coverage, at which the transition from 2D to 3D growth  occurs, are presented in the literature. For example, an abrupt increase in hut density at the coverage of 3.16\,ML was detected  for Ge deposition at {300\,\textcelsius} and 0.06\,ML/min.\cite{Jesson_Kastner_Voigt} A detailed phase diagram of the Ge film on Si(001) derived from experiments carried out by recording diffractometry of reflected high-energy electrons (RHEED) gave the coverages corresponding to the ``2D-to-hut'' transition from $\sim 2.5$ to $\sim 3$\,ML for the growth temperature interval from 300 to 400{\,\textcelsius} (and different values for different temperatures).\cite{Nikiforov_Gi-Si_RHEED} Photoluminescence (PL) study of Ge huts deposited at the temperature of 360{\,\textcelsius} showed that evolution from ``quantum-well-like'' (attributed to WL) to ``quantum-dot-like'' (attributed to Ge huts) emission occurred at a coverage of $\sim 4.7$\,ML in  PL spectra obtained at 8\,K.\cite{PL_ultrasmall_Ge} Hut formation studied by high resolution low-energy electron diffraction (LEED) and surface-stress-induced optical deflection  evidenced that at deposition temperature of 500{\,\textcelsius} hut formation suddenly set in at a coverage of 3.5\,ML.\cite{LEED_Ge-nucleation} And finally, for theoretical studies the WL thickness and consequently the hut formation coverage is usually assumed  to equal 3\,ML.\cite{[{See, e.\,g., }]hut_stability} As it is seen from the above examples,  there is no unambiguous information presently about the coverage at which huts arise or, more accurately, about the thickness of the WL $M\times N$  patch on which a cluster nucleate during Ge deposition. STM studies show the WL thickness to equal 3\,ML only on the average: $M\times N$ patches have slightly different thicknesses (\textpm\,1\,ML) around this value.\cite{Iwawaki_initial,classification,CMOS-compatible-EMRS,Hut_nucleation,atomic_structure}
In this article, we determine by means of high resolution STM an accurate value of a Ge coverage at which hut array nucleate at 360{\,\textcelsius}. In addition, we investigate the early stage of the array evolution and explore the patch surface reconstruction as a result of hut appearance. 

\begin{figure*}
\includegraphics[scale=1.1]{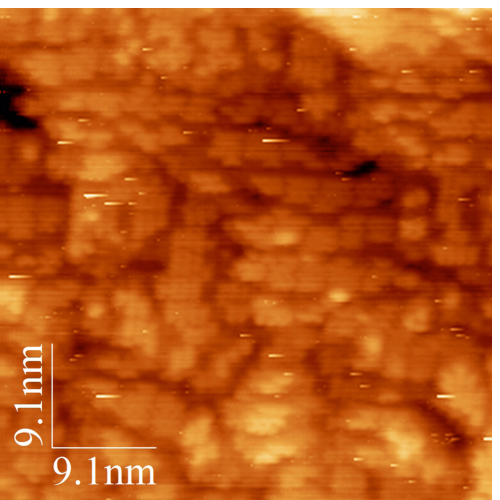}(a)
\includegraphics[scale=1.1]{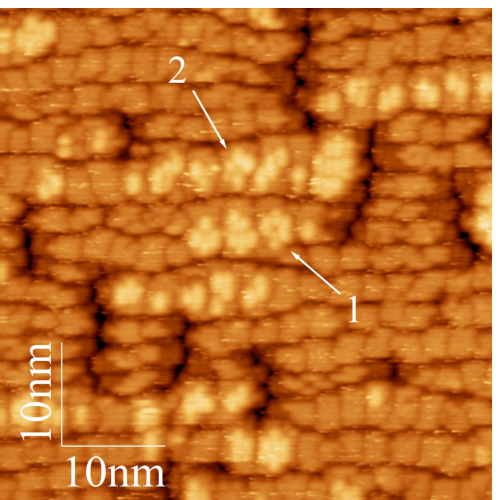}(b)
\includegraphics[scale=1.1]{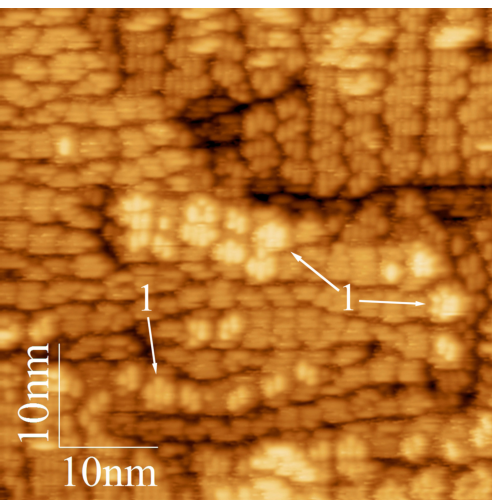}(c)
\includegraphics[scale=1.1]{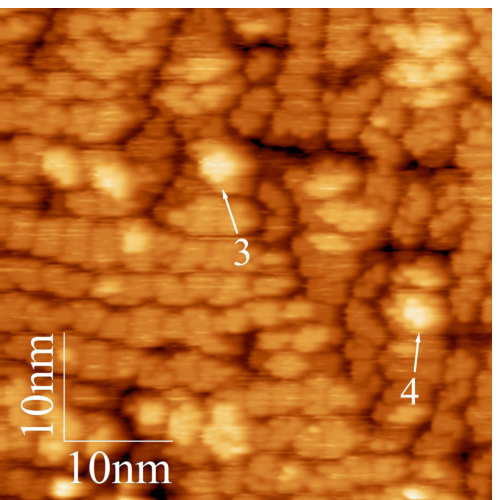}(d)
\caption{\label{fig:4,4A_5,1A}(Color online)  
STM images of  Ge  wetting layer on Si(001): 
(a) $h_{\rm Ge}=4.4$~\r{A}  ($U_{\rm s}=-1.86$~V, $I_{\rm t}=100$~pA), neither hut clusters nor their nuclei are observed;  
(b) $h_{\rm Ge}=5.1$~\r{A},  $U_{\rm s}=+1.73$~V, $I_{\rm t}=150$~pA; 
(c) $U_{\rm s}=+1.80$~V, $I_{\rm t}=100$~pA; 
(d) $U_{\rm s}=+2.00$~V, $I_{\rm t}=100$~pA. 
Examples of characteristic features are numbered as follows:  nuclei of pyramids (1) and  wedges  (2) [1\,ML high over WL patchs, Fig.~\ref{fig:nuclei}],\cite{Hut_nucleation,CMOS-compatible-EMRS}  small pyramids (3) and wedges (4) [2\,ML high over WL patchs]\cite{Hut_nucleation,CMOS-compatible-EMRS,classification,atomic_structure}}.
\end{figure*}

Our approach to the problem is simple. Recently we described two characteristic formations composed by epitaxially oriented Ge dimer pairs and  chains of four dimers on the WL patches which were interpreted as two types of hut nuclei: an individual type for each species of huts---pyramids or wedges (Fig.~\ref{fig:nuclei}).\cite{Hut_nucleation} Being aware of the shapes of the nuclei we can determine by STM a coverage ($h_{\rm Ge}$) at which the first  generation of nuclei emerge on WL. Then we can trace the evolution of an array and WL until huts with pronounced shapes and faceting form.

\section{Experimental}

Experiments were carried out using a UHV MBE chamber (residual gas pressure $P\sim 10^{-11}$\,Torr) coupled with STM  ($P\sim 10^{-10}$\,Torr).\cite{classification,CMOS-compatible-EMRS} Substrates  were 8$\times$8 mm$^{2}$ squares cut from the specially treated commercial B-doped    CZ Si$(100)$ wafers ($p$-type,  $\rho\,= 12~\rm\Omega\,$cm). Ge was deposited on the  clean Si(001) surface\cite{stm-rheed-EMRS} from a source with the electron beam evaporation. The  deposition rate was $\sim 0.15$\,\r{A}/s; $h_{\rm Ge}$ was varied from 3 to 6\,\r{A} for different samples;  $h_{\rm Ge}$ and $dh_{\rm Ge}/dt$ were controlled by the Inficon Leybold-Heraeus  XTC\,751-001-G1 thin film deposition thickness and rate controller with a quartz sensor.
The substrate temperature 
was 360{\,\textcelsius}; the pressure in the MBE chamber did not exceed $10^{-9}$\,Torr during Ge deposition. The rate of the sample cooling down to the room temperature was approximately 0.4{\,\textcelsius}/s. After  cooling, the samples were moved  into the STM chamber. The images were obtained in the constant tunneling current ($I_{\rm t}$) mode at the room temperature. The STM tip was zero-biased while the sample was positively or negatively biased ($U_{\rm s}$) for empty or filled states imaging.
STM images were processed using the WSxM software.\cite{WSxM}
A detailed description of the experimental procedures can be found in Ref.~\onlinecite{classification,CMOS-compatible-EMRS}.

\section{Data and Discussion}

\begin{figure*}
\includegraphics[scale=1.1]{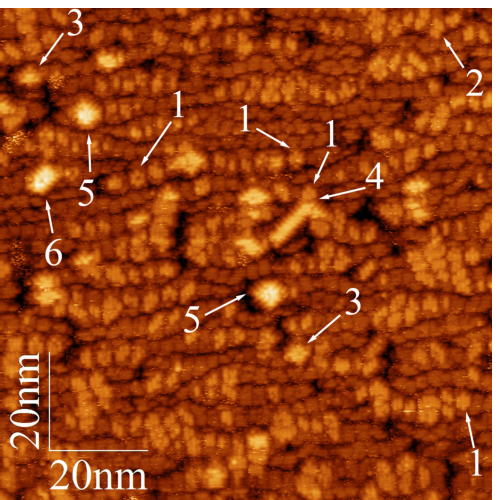}(a)
\includegraphics[scale=1.1]{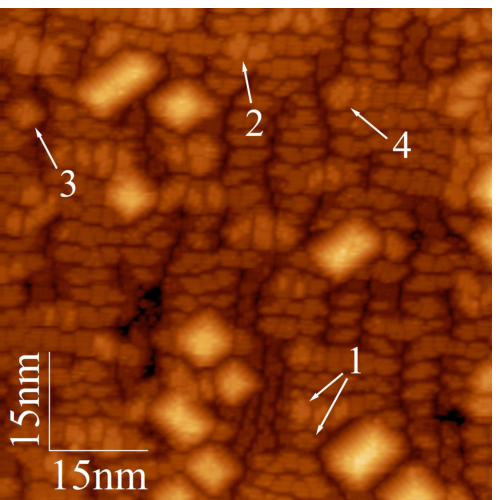}(b)
\caption{\label{fig:5,4A_6A}(Color online)  
STM images of  Ge wetting layer on Si(001),  
(a) $h_{\rm Ge}$ is 5.4\,\r{A}  ($U_{\rm s}=+1.80$~V, $I_{\rm t}=100$~pA)  
and 
(b) 6.0~\r{A}   ($U_{\rm s}=+2.50$~V, $I_{\rm t}=80$~pA).  
The numbering is as follows:  
 (1) to (4) are the same as in Fig.\,\ref{fig:4,4A_5,1A};  (5) and (6) are 3\,ML high pyramids (5)   and  wedges (6).}
\end{figure*}

\begin{figure*}
\includegraphics[scale=1.145]{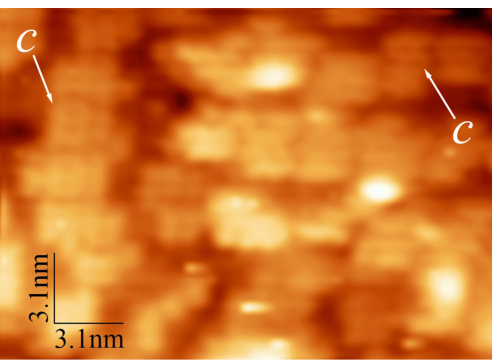}(a)
\includegraphics[scale=1.2]{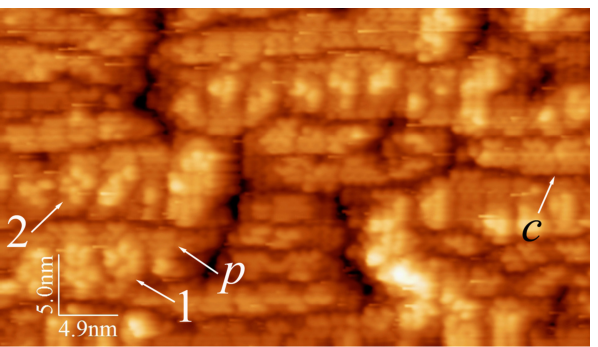}(b)
\includegraphics[scale=1.2]{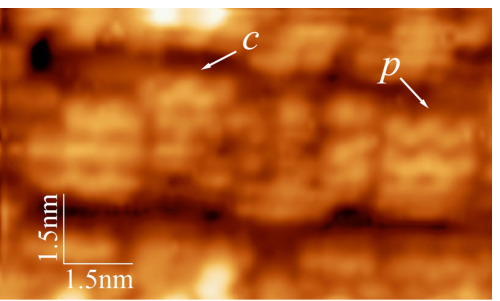}(c)
\includegraphics[scale=1.007]{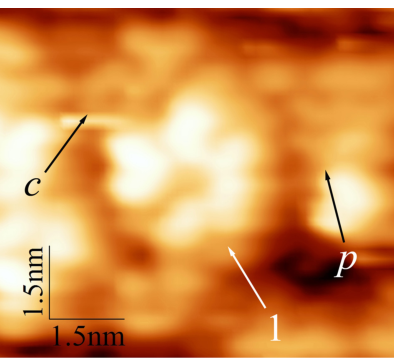}(d)
\caption{\label{fig:WL_patches}(Color online)  
STM images of  Ge wetting layer on Si(001): the ordinary $c(4\times 2)$~$(c)$  and $p(2\times 2)$~$(p)$ reconstructions within the $M\times N$ patches are often observed simultaneously,  
(a) $h_{\rm Ge}=4.4$~\r{A}, $U_{\rm s}=-1.86$~V, $I_{\rm t}=100$~pA, 
 only the $c(4\times 2)$ structure is resolved; 
(b)~$h_{\rm Ge}=5.1$~\r{A}, $U_{\rm s}=-3.78$~V, $I_{\rm t}=100$~pA, 
 both $c(4\times 2)$  and $p(2\times 2)$ structures are revealed as well as  nuclei of a pyramid (1) and a wedge (2); 
(c) $h_{\rm Ge}=6.0$~\r{A}, $U_{\rm s}=+1.80$~V, $I_{\rm t}=80$~pA, both $c(4\times 2)$  and $p(2\times 2)$ reconstructions are well resolved; 
(d) $h_{\rm Ge}=5.1$~\r{A}, $U_{\rm s}=-3.78$~V, $I_{\rm t}=100$~pA, a pyramid nucleus  on the $c(4\times 2)$ reconstructed patch with the adjacent  $p(2\times 2)$  reconstructed patch.}
\end{figure*}

The obtained experimental data are as follows. 
Fig.\,\ref{fig:4,4A_5,1A}(a) demonstrates a typical STM micrograph of the  ($M\times N$)-patched WL ($h_{\rm Ge}=4.4$~\r{A}, $\sim 3.1$\,ML). This image does not demonstrate any feature which might be recognized as a hut nucleus (Fig.~\ref{fig:nuclei}).\cite{Hut_nucleation} Such features first arise at the  coverages of $\sim 5$\,\r{A}: they are clearly seen in the images (b) to (d), which demonstrate a moment when the array have just nucleated ($h_{\rm Ge}=5.1$\,\r{A}, $\sim 3.6$\,ML). However, we succeeded to find minute pyramid and wedge at this $h_{\rm Ge}$ (Fig.~\ref{fig:4,4A_5,1A}(d))---both as small as 2\,ML over the patch surface (we measure cluster heighs from  patch tops)---which indicate that hut nucleation had started a little earlier. 

It can be concluded from these observations that hut arrays nucleate at a coverage of $\sim 5.1$\,\r{A} ($\sim 3.6$\,ML) when approximately a half of patches are as thick as 4\,ML. We can suppose then that huts nucleate on those patches whose thickness reaches (or even exceeds) 4\,ML. 

The hut nucleation goes on during  further evolution of the array. Fig.\,\ref{fig:5,4A_6A}
illustrates this process. An array shown in Fig.\,\ref{fig:5,4A_6A}(a) ($h_{\rm Ge} = 5.4$\,\r{A}, $\sim 3.9$\,ML) consists of 1-ML nuclei, 2-ML and 3-ML pyramids and wedges. 
Fig.~\ref{fig:5,4A_6A}(b) ($h_{\rm Ge} = 6.0$~\r{A}, $\sim 4.3$\,ML) demonstrates the simultaneous presence of nuclei  and 2-ML huts  with the growing much higher clusters. So, hut arrays initially evolve with increasing  $h_{\rm Ge}$
by   concurrent growth of available clusters and nucleation of new ones resulting in progressive rise of  hut number  density.

Our data are in very good agreement with the results reported in Ref.\,\onlinecite{LEED_Ge-nucleation}, except for the growth temperature.

Evolution of WL patches during MBE is illustrated by Fig.~\ref{fig:WL_patches}. In full agreement with the data of Ref.\,\onlinecite{Iwawaki_initial}, both $c(4\times 2)$ and $p(2\times 2)$ reconstructions are observed on tops of the $M\times N$ patches in all images except for the image Fig.~\ref{fig:WL_patches}(a) ($h_{\rm Ge}=4.4$~\r{A}) in which only the $c(4\times 2)$ structure is recognized. Formation of a hut nucleus on a patch reconstructs its surface; a new formation changes the topmost layer structure to that specific for a particular type of nuclei, in the present case, to the structure of the pyramidal hut nucleus (Fig.~\ref{fig:WL_patches}(d)). However the residual $c(4\times 2)$ structure  still remains on the lower terrace of the patch. At the same time, the $p(2\times 2)$ structure stays on the top of the adjacent patch.

It is necessary to remark here that the nuclei are always observed to arise on sufficiently large WL patches. There must be enough room for a nucleus on a single patch. A nucleus cannot be housed on more than one patch. Thus, cluster nucleation is impossible on little (too narrow or short) patches (Figs.~\ref{fig:nuclei} and~\ref{fig:WL_patches}(d)).\cite{classification,Hut_nucleation,atomic_structure}

Note also that both types of nuclei emerge at the same moment of the MBE growth. It means that they are degenerate by the formation energy. 
Obvious consideration resulting in this conclusion is following: both types of the hut nuclei arise at the same WL thickness (``moment''), hence, at the same WL stress to relief it. So, they appear at the same strain energy (and with equal likelihoods, see Refs.~\onlinecite{classification,Hut_nucleation}). If they had different formation energies they would appear at different WL thicknesses. The first of the types of huts, which nucleates on the surface, releases the stress.  The second one never appears therefore. Hence, they can appear only simultaneously. And their formation energies, as it follows from our observations, can only be equal.  Calculations supporting (or refuting) our reasoning are desirable, however, for explanation of why two structures  different in symmetry have equal energies and probabilities of formation.
 Until then, an issue of a reason which makes   two different structures arise, rather than one,   to  relief the WL strain  remains open, however.

As of now, we can only propose a very preliminary interpretation of the observed simultaneous appearance of the two kinds of nuclei on WL  at the patch thickness of 4\,ML. The explanation is based on  modeling of Ge cluster formation energy performed in Ref.~\onlinecite{Domes_first}. The authors of  Ref.~\onlinecite{Domes_first} explore Ge island nucleation during MBE  at much higher temperatures than those applied in this work, therefore  theoretical results of Ref.~\onlinecite{Domes_first} describe the experimental data obtained for the  case of the high-temperature growth mode, which differs considerably from the low-temperature one.\cite{classification} However, the modeling could also apply  for the low-temperature growth. The case is that according to Ref.~\onlinecite{Domes_first}, flat Ge islands---nuclei and small huts---likely occur on WL because of an energy benefit which arises in exposing the compressed \{105\} facets, rather than in relaxing the volumetric elastic energy, as it takes place in the usual Stranski-Krastanov mechanism. At low temperatures, this effect may stabilize clusters, however preventing their further ripening (this agrees with our observations presented recently in Ref.~\onlinecite{classification}). If this is the case, the actual volumetric form of clusters really does not matter very much in their formation energy, and nucleation probabilities (and energies) of the \{105\} faceted pyramids or wedges appear to be close in spite of difference in their symmetries. 

This model may also be useful for explaining swift elongation of wedges as well as gradual extinction of pyramids during low temperature MBE.\cite{classification} Notice also that the degeneracy of wedge facets is likely removed by a vacancy-type defect which is always present on each triangular facet of a wedge-like hut.\cite{Hut_nucleation,CMOS-compatible-EMRS}

\section{Conclusion}

In summary, the following conclusions are made. 
At 360\,\textcelsius, nuclei of Ge pyramids and wedges are observed on the  wetting layer $M\times N$ patches starting from the coverage of about $5.1$\,\r{A} ($\sim 3.6$\,ML). This suggests that huts nucleate on patches of   4\,ML thick; the formation energies  of both types of nuclei  are equal. Further development of hut arrays consists in simultaneous growth of the formerly emerged clusters and nucleation of new ones resulting in gradual rise of  hut number  density with increasing $h_{\rm Ge}$. Huts nucleate reconstructing the patch surface from the usual $c(4\times 2)$ or $p(2\times 2)$ structure to one of two recently described more complicated  formations composed by epitaxially oriented Ge dimer pairs and  chains of four dimers.

\begin{acknowledgments}

This work was supported  by
the Ministry of Education and Science of the Russian Federation, the State Contract no.\,14.740.11.0069.

\end{acknowledgments}

\bibliography{Array_nucleation}

\end{document}